\documentclass{svmult}

\usepackage{graphicx}
\usepackage{xspace}

\newcommand{\dd}{\ensuremath{\mathrm{d}}}
\newcommand{\refsec}[1]{Section~\ref{#1}}
\newcommand{\reffig}[1]{Fig.~\ref{#1}}
\newcommand{\refeq}[1]{equation~(\ref{#1})}

\newcommand{\sn}{\hbox{$\mathcal{S/N}$}\xspace}
\newcommand{\snt}{\hbox{$\mathcal{(S/N)_\mathrm{T}}$}\xspace}
\newcommand{\snb}{\hbox{$\mathcal{(S/N)_\mathrm{bin}}$}\xspace}
\newcommand{\apj}{ApJ\xspace}           
\newcommand{\apjl}{ApJ\xspace}           
\newcommand{\mnras}{MNRAS\xspace}       

\newcommand{\aap}{A\&A\xspace}
\newcommand{\aaps}{A\&AS\xspace}

\newcommand{\aj}{AJ\xspace}
\newcommand{\pasp}{PASP\xspace}

\newcommand{\pasj}{PASJ\xspace} 

\begin{document}

\title*{Voronoi binning: Optimal adaptive tessellations of multi-dimensional data}
\titlerunning{Adaptive Voronoi binning}
\author{Michele Cappellari\inst{1}}
\institute{$^1$ Sub-department of Astrophysics, University of Oxford, Denys Wilkinson Building, Keble Road, Oxford, OX1 3RH, England \texttt{cappellari@astro.ox.ac.uk}}

\maketitle

\begin{abstract}
We review the concepts of the Voronoi binning technique \cite{2003MNRAS.342..345C}, which optimally solves the problem of preserving the maximum spatial resolution of general two-dimensional data, given a constraint on the minimum signal-to-noise ratio (\sn). This is achieved by partitioning the data in an adaptive fashion using a Voronoi tessellation with nearly hexagonal lattice. We review astrophysical applications of the method to X-ray data, integral-field spectroscopy, Fabry-Perot interferometry, N-body simulations, standard images and other regularly or irregularly sampled data. Voronoi binning, unlike adaptive smoothing, produces maps where the noise in the data can be visually assessed and spurious artifacts can be recognized. The method can be used to bin data according to any general criterion and not just \sn. It can be applied to higher dimensions and it can be used to generate optimal adaptive meshes for numerical simulations.
\end{abstract}

\section{Introduction}

In scientific research one often wants to measure a physical quantity at a certain set of positions in two (or more) dimensions. The quantity under exam can vary by many orders of magnitude between the different locations. When the quantity represents the intensity of a signal $\mathcal{S}$, its ratio with the corresponding noise $\mathcal{N}$ will correspondingly vary. In general one wants to perform subsequent analysis on the recorded signal, for example to extract derived quantities, to compare the observations with model predictions or to better visualize and interpret the observed signal. In these cases it is useful to adaptively bin (or partition, or segment) the data {\em to preserve the maximum spatial resolution, given a constraint on the required minimum \sn}.

Adaptive binning should not be improperly called or confused with adaptive smoothing \cite{1986desd.book.....S}, which consists of correlating neighbouring information \cite{1991AJ....102.1581B,1994AJ....108..514M,1996ApJ...461..622H,
1998A&AS..128..397S,2006MNRAS.368...65E}.
Smoothing is not ideal when an accurate error analysis is needed, it does not allow for a visual assessment of the noise level in the data and it makes spurious artifacts difficult to recognize. A comparison between Voronoi binning and adaptive smoothing is presented in \cite{2006MNRAS.368..497D}.

\section{Why is it useful to bin the data?}

\begin{figure}[tbp]
\centering
\includegraphics[width=\columnwidth]{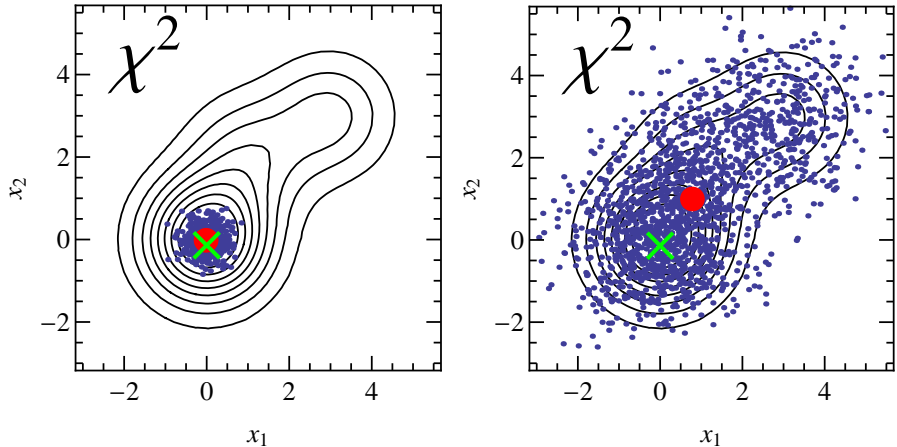}
\caption{$\chi^2$ contour plots for an idealized physical model. The blue dots represent the measurements for the two model parameters $(x_1,x_2)$ for different random realizations of the data. The large red circle is the average of the realizations, while the green cross marks the minimum of $\chi^2$.
{\em Left panel:} At high-\sn\ all measurements lie close to the minimum, where the $\chi^2$ is well approximated by a quadratic form and the average of the realizations agrees with the true minimum. {\em Right panel:} At low-\sn\ the measurements can explore higher levels of $\chi^2$, where the contours become asymmetric. In this case the average solution will be biased.}
\label{fig:chi2}
\end{figure}

The main theoretical motivation for binning is to reduce the bias that can be introduced, in the low-\sn\ regime, when fitting a general nonlinear physical model to the data. The $\chi^2$, which describe the agreement between data and model, is in general approximated by a quadratic form near the minimum, but far from it may become asymmetric (unless the model is fully linear, which is rarely the case in practice). An illustration of the situation is presented in \reffig{fig:chi2}: When the data have high \sn\ the perturbed solutions stay close to the minimum and the average solution is unbiased. But at low \sn\ the perturbed solutions can explore higher levels of $\chi^2$, where the contours become asymmetric. In this case the average solution does {\em not} agree with the true minimum and is biased. Binning the data to increase the \sn\ and fitting the model to them can still provide an unbiased solution to this problem.

\section{How to define the optimal binning?}
\label{sec:optimal_binning}

\begin{figure}[tbp]
\centering
\includegraphics[width=0.49\columnwidth]{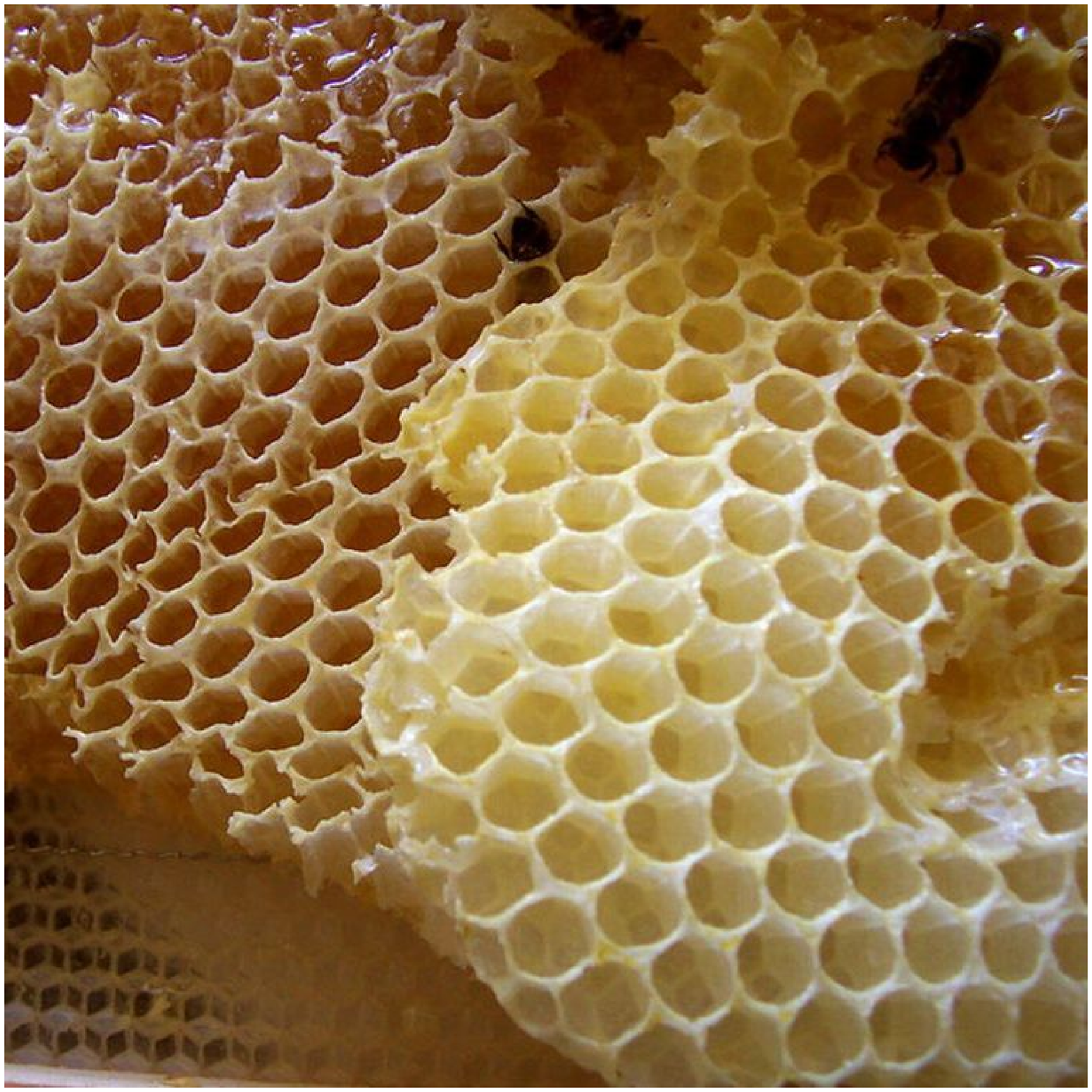}
\includegraphics[width=0.49\columnwidth]{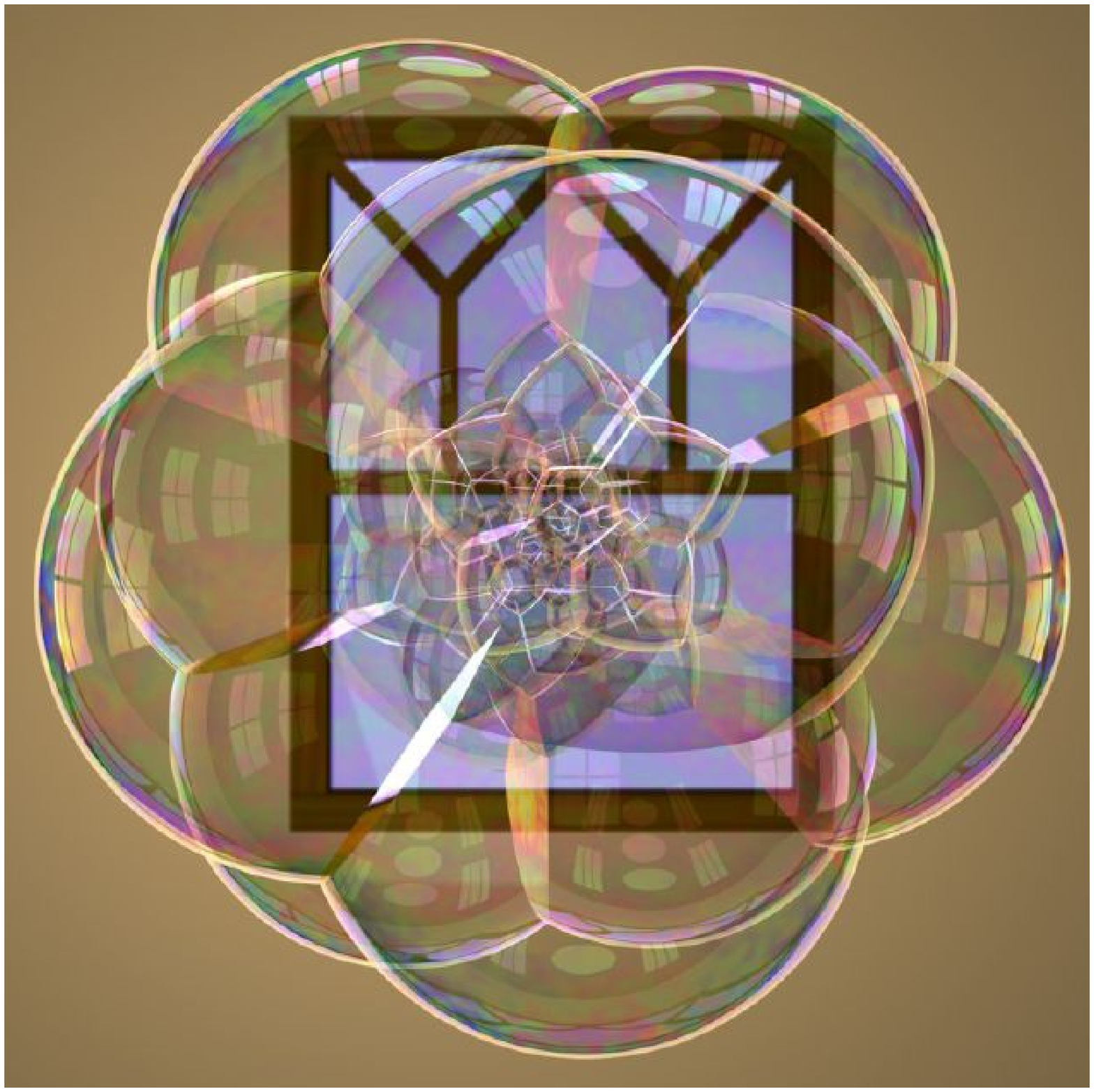}
\caption{{\em Left panel:} A honeycomb consists of a lattice of hexagonal cells. {\em Right panel:} A simulated cluster of bubbles (from \protect\cite{sullivan1991generating}). The bubbles adapt their size to the enclosed amount of air, while keeping a compact (minimum surface) configuration. This constitutes an ideal physical model for an adaptive binning method in two and three dimensions.}
\label{fig:bubbles}
\end{figure}

Earlier attempts to bin two dimensional data relied on recursive methods. A classic one is the Quadtree \cite{samet1984quadtree} which recursively partitions a region into axis-aligned squares. A variation was proposed by \cite{2001MNRAS.325..178S} and applied to the adaptive binning of X-ray images. The limitation of recursive methods is that they can only employ predefined sizes for the bins, which are powers of a basic element. This necessarily produces discontinuities in the \sn\ and artifacts. To improve on recursive methods \cite{2003MNRAS.342..345C} identified three criteria for the ideal adaptive binning:
\begin{description}
  \item[\bf Topological:] The data have to be partitioned without holes or overlapping bins: One wants to use all available data, without duplications;
  \item[\bf Morphological:] The bins should be as compact or round as possible: One wants to preserve the maximum spatial resolution;
  \item[\bf Uniformity:] The scatter in \sn\ should be minimized: One wants to increase the \sn\ to the required target level \snt, but not above that, as that would degrade the spatial resolution.
\end{description}

The criteria are quite general. They define the goal to partition the plane in a set of nearly circular\footnote{An alternative is discussed in the Contour Binning method \cite{2006MNRAS.371..829S}.} and maximally-packed regions enclosing equal amount of a certain quantity. This generality is one of the reasons why the criteria happen to be satisfied in a number of physical phenomena in Nature. An optimal partition of the plane is naturally produced in honeycombs, which consist of equal-size nearly hexagonal cells (\reffig{fig:bubbles}). Similar partitions with hexagonal lattice are naturally generated in convective cells, for example in the granulation of the solar photosphere. In these cases however the physical conditions that drive the plane partition are spatially constant and the cells are all equal. A beautiful example where an adaptive partition of space is naturally produced by an optimality criterion, in that case the minimization of the surface, is provided by the configuration of a cluster of soap bubbles (\reffig{fig:bubbles}). This phenomenon constitutes an ideal physical model for the optimal binning method in two or three dimensions.

\section{The Voronoi binning method}
\label{sec:voronoi_method}

To simplify the terminology in what follows we call `pixel' an individual data element. This is not just the pixel of an uniformly sampled image but represents a generic quantity which is measured at a certain location identified by two (or more) coordinates. In applications to integral-field spectroscopy, for example, a pixel refers to a full spectrum measured at one sky location.

The Voronoi binning method consists of two separate stages: (i) The bin-accretion stage, which generates an initial set of bins satisfying the criteria of \refsec{sec:optimal_binning} and (ii) an optional regularization stage, which tries to improve the binning produced at the previous stage. Here we summarize the main ideas of the method. A detailed description of the technique, including essential details, is given in \cite{2003MNRAS.342..345C}. A reference implementation is available online\footnote{http://purl.org/cappellari/idl}.

\subsection{The bin-accretion stage}
\label{sec:bin_accretion}

\begin{figure}[tbp]
\centering
\includegraphics[width=0.50\columnwidth]{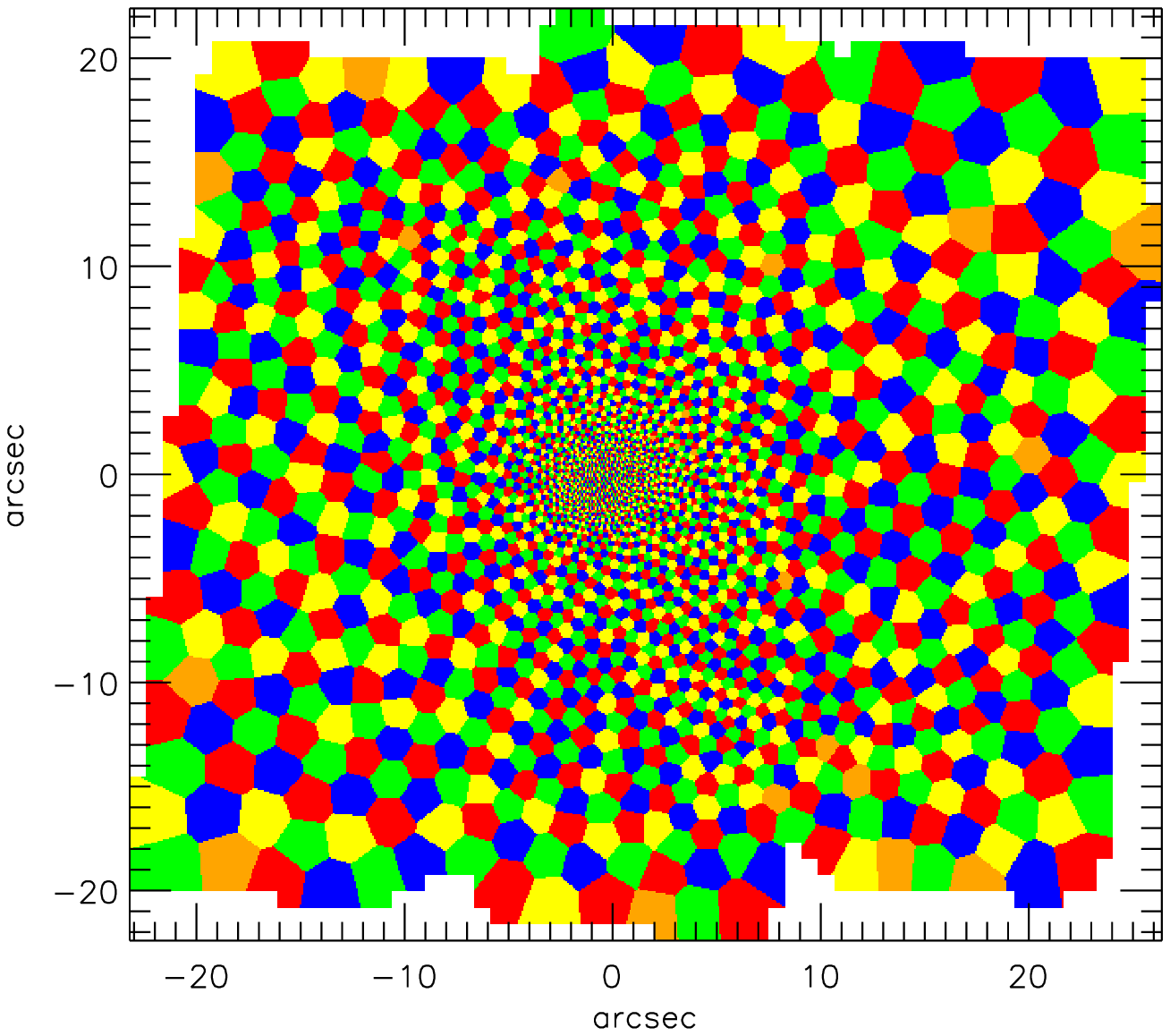}
\includegraphics[width=0.49\columnwidth]{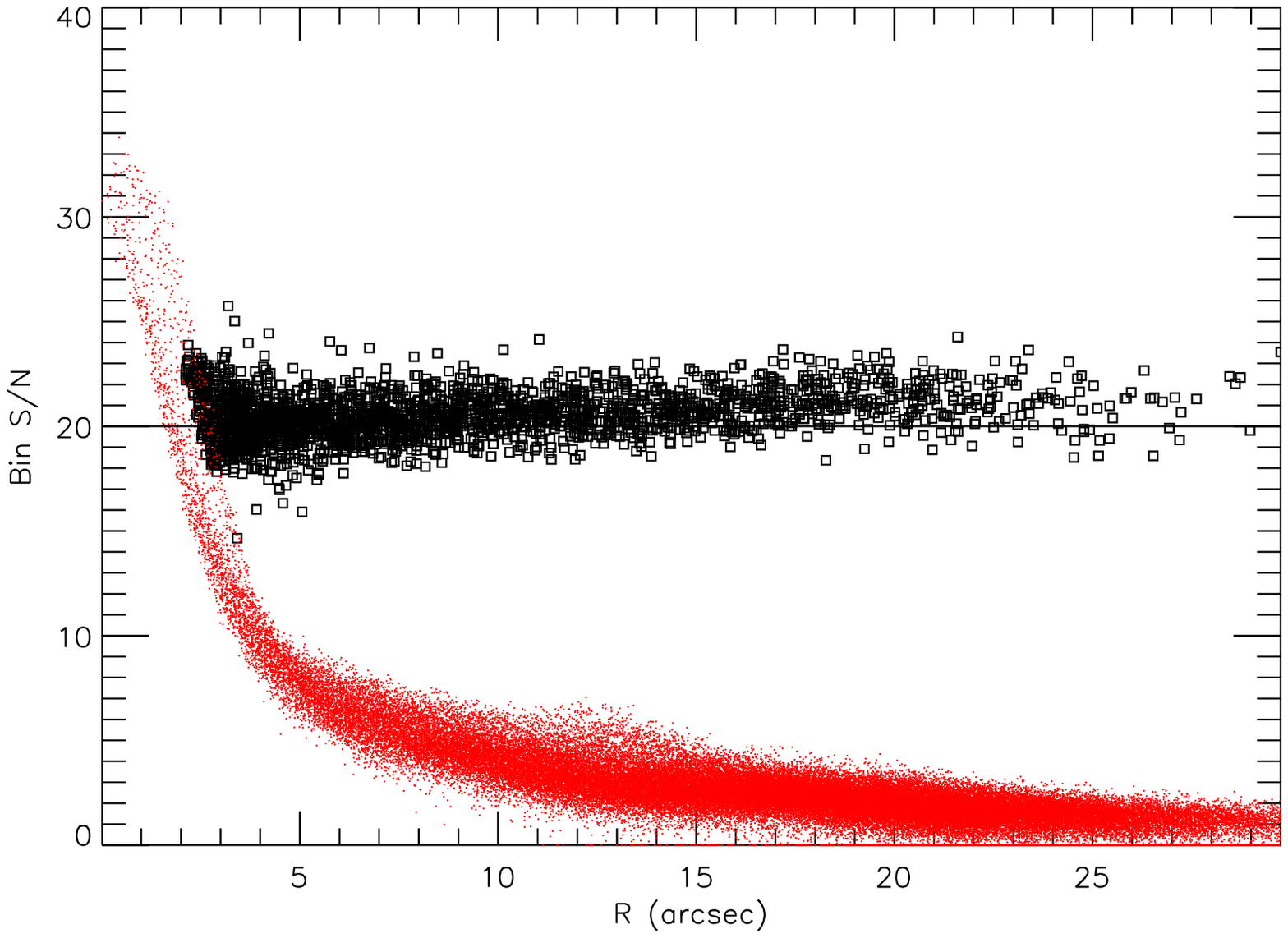}
\caption{The bin-accretion stage. {\em Left panel:} Tessellation generated by the bin-accretion stage of the Voronoi binning method for an image with about 350,000 pixels, grouped into 3000 bins. The size of the bins adapts to the \sn\ while keeping a compact shape. {\em Right panel:} The \sn\ of the original pixels is shown with the red points and the \snb\ of the bins with the open squares. The target $\snt=20$ is indicated by the solid line.}
\label{fig:bins_accretion}
\end{figure}

The Voronoi binning method starts by trying to generalize the process used to bin one-dimensional data (e.g. \cite{1994MNRAS.270..271V}). In one dimension one starts from the pixel with the highest \sn\ and keeps `accreting' neighbouring pixels, until the target \snt\ is reached. In two (or more) dimensions one can still accrete pixels that are closest to the current bin, however there will be many adjacent pixels and one has to decide which one to accrete next. Moreover the \sn\ criterion is not sufficient to decide when to start a new bin. The bin shape also plays a role as the bin may become too elongated e.g.\ when the edge of the data is reached. The bin-accretion stage of the Voronoi binning algorithm solves these problems with two simple choices:
\begin{enumerate}
  \item The next pixel to accrete into the current bin is the closest to its centroid;
  \item The accretion of new pixels into a bin continues as long as the criteria for the optimal binning of \refsec{sec:optimal_binning} are satisfied.
\end{enumerate}
This process continues, pixel-by-pixel, until all pixels in the image have been processed. At the end of the process however, some pixels will generally remain unbinned and a few bins will not have reached the desired \snt. The method computes the geometric centroid $\mathbf{z}_j$ of all good bins and these are used as generators for a Voronoi tessellation (VT; see \cite{Okabe2000}), which partitions the data into convex bins $V_j$ (\emph{Voronoi regions}): The pixel $\mathbf{x}_i$ is assigned to the bin that minimizes the distance $d_i=\|\mathbf{z}_j-\mathbf{x}_i\|$.

An example of binning produced by the bin-accretion stage of the method is shown in \reffig{fig:bins_accretion}. The bins tend to the desired hexagonal lattice, but their size adapts to the \sn\ in such a way that they have nearly the same \sn. A possible choice to estimate the \sn\ of a bin during the accretion is to define
\begin{equation}
    \snb = \frac{\sum_i \mathcal{S}_i}{\sqrt{\sum_i \mathcal{N}_i^2}},
    \label{eq:sn}
\end{equation}
where the summation is done over the pixels in that bin. {\em However the bin-accretion algorithm is completely independent on the way in which \snb\ is estimated!}

In some applications the noise of the individual pixels may be correlated, so that \refeq{eq:sn} is not accurate. Or it may be possible to determine \snb\ only when it is sufficiently high. For example when binning clumpy gas in a galaxy the $\mathcal{S}_i$ in many pixels may be undetected before binning. Alternatively one may want to bin the data not according to \snb, but according to the errors in an extracted quantity (e.g.\ kinematics) from the bin. All these situations can be trivially included in the bin-accretion algorithm by replacing \refeq{eq:sn} with a procedure to estimate \snb, or any other quantity of interest, during the bin accretion.

Due to its generality of estimating the \sn\ in a bin, and due to its ability to produce small scatter in the \sn, also in the few-pixels regime, this bin-accretion stage may be all that one needs to Voronoi bin the data. Optionally however one may start from the binning generated by this stage and further improve it via some iterative process. We call this the bin-regularization stage, which is the subject of the next section.

\subsection{The bin-regularization stage}

In the two important cases where (i) the noise is approximately Poissonian (shot noise), or (ii) when the pixels in each bin are optimally weighted with weights $w=\mathcal{S}_j/\mathcal{N}^2_j$ to maximize \snb\ \cite{1986PASP...98.1220R,1986PASP...98..609H}, then \refeq{eq:sn} reduces to $(\mathcal{S/N})_\mathrm{bin}^2=\sum_i (\sn)^2_i$. One can then define a density distribution $\rho(\mathbf{y})=(\sn)^2(\mathbf{y})$ such that the problem of binning to a constant \sn\ reduces to that of obtaining a tessellation enclosing equal mass according to $\rho$. In that case a \emph{Centroidal Voronoi Tessellation} (CVT) can be used to regularize the binning provided by the bin-accretion stage.

Given a density function $\rho(\mathbf{r})$, a CVT is a special class of VT where the generators $\mathbf{z}_{j}$ happen to coincide with the mass centroids
\begin{equation}
  \mathbf{z}_{j}^{\ast} =
  \frac{\int_{V_{j}} \mathbf{y}\rho(\mathbf{y})\,\dd\mathbf{y}}
  {\int_{V_{j}} \rho(\mathbf{y})\,\dd\mathbf{y}}
  \label{eq:centroid}
\end{equation}
of the corresponding Voronoi regions $V_j$ (see review \cite{du1999centroidal}).
A striking characteristic of the CVT in the two-dimensional case is its
ability to partition a region into bins whose size varies as a
function of the underlying density distribution, but whose shape tends
asymptotically to an hexagonal lattice for a large number of bins.
Another nice feature of the CVT is that a simple algorithm exists for its
practical computation, Lloyd's method \cite{lloyd1982least}, which simply consists of a repeated application of \refeq{eq:centroid}. In the statistical literature (e.g. \cite{hartigan1975clustering}) discrete CVTs are related to optimal $k$-means clusters, where Voronoi regions and centroids are referred to as clusters and cluster centers, respectively.

Although CVT bins are naturally smaller where the density is higher,
the mass $M_j=A_j\rho_j$ enclosed in a bin, where $A_j$ is the bin area, is not constant. However \cite{2003MNRAS.342..345C} demonstrated that if the CVT is constructed for the density $\rho'=\rho^2$, then Lloyd's algorithm converges toward an equal-mass CVT according to $\rho$.

An alternative to the regularization stage using the equal-mass CVT has been proposed by \cite{2006MNRAS.368..497D}. It consists of using a Weighted Voronoi Tessellation (WVT; \cite{møller1994lectures}), which is a generalization of a VT in which bins are not uniquely defined by the generators, but also have an associated scale $\delta_j$. The Lloyd's iteration of the CVT scheme is modified as follows: (i) A pixel with coordinates $\mathbf{x}_i$ is assigned to the bin that minimizes the {\em scaled }distance $d_i=\|\mathbf{z}_j-\mathbf{x}_i\|/\delta_j$, producing a WVT (where $\delta_j=1$ reduces to the standard VT); (ii) The density is set to a constant $\rho(\mathbf{y})=1$, so that $\mathbf{z}_{j}^{\ast}$ in \refeq{eq:centroid} becomes the geometric centroid; (iii) The scale of each bin during Lloyd's iteration is updated as $\delta_j=\sqrt{A_j/\snb}$. Clearly a binning with constant \sn\ across the field is a stable fixed point of this iteration scheme, as it satisfies the relation $\delta_j/\delta_k=\sqrt{A_j/A_k}$ (see \cite{2006MNRAS.368..497D} for details). The advantage of regularizing the bins using a WVT instead of a CVT, is that the former, like the bin-accretion algorithm, allows for a completely general form of the noise. The bins also tend to be more compact at the  edge of the field. A generally minor disadvantage is that the resulting binning is not a standard VT: The scales $\delta_j$ are required to describe the bins, together with the generators $\mathbf{z}_j$ and the bins are not necessarily convex.

As a practical advice, in most situations the bin-accretion algorithm of \refsec{sec:bin_accretion} produces a binning that already closely satisfies the optimal criteria of \refsec{sec:optimal_binning} (\reffig{fig:bins_accretion}). If one wants the binning to be described by a VT, and if the noise is approximately Poissonian (or the pixels are optimally weighted), one can apply the CVT regularization stage. In the other cases one may just skip the regularization stage or apply a WVT regularization. All these options are available in the reference implementation\footnotemark[2] and are quick to test on the data at hand. An implementation adapted by \cite{2006MNRAS.368..497D}, including specific examples of usage on X-ray data, is also available online\footnote{http://www.phy.ohiou.edu/$\sim$diehl/WVT/}.

The size of the equal-mass CVT can be adapted to any desired density distribution while still preserving the nearly hexagonal lattice. The same is true for the WVT. This makes the algorithm an excellent mesh generator if one considers that the dual of a Voronoi tessellation with nearly hexagonal bins is a Delaunay triangulation with nearly equilateral triangles \cite{2003MNRAS.342..345C,du2002grid}. A recent astrophysical illustration of the power of this general idea is given by the mesh regularization scheme in the AREPO cosmological simulation code \cite{2009MNRAS.tmp.1655S}.

\begin{figure}[tbp]
\centering
\includegraphics[width=0.49\columnwidth]{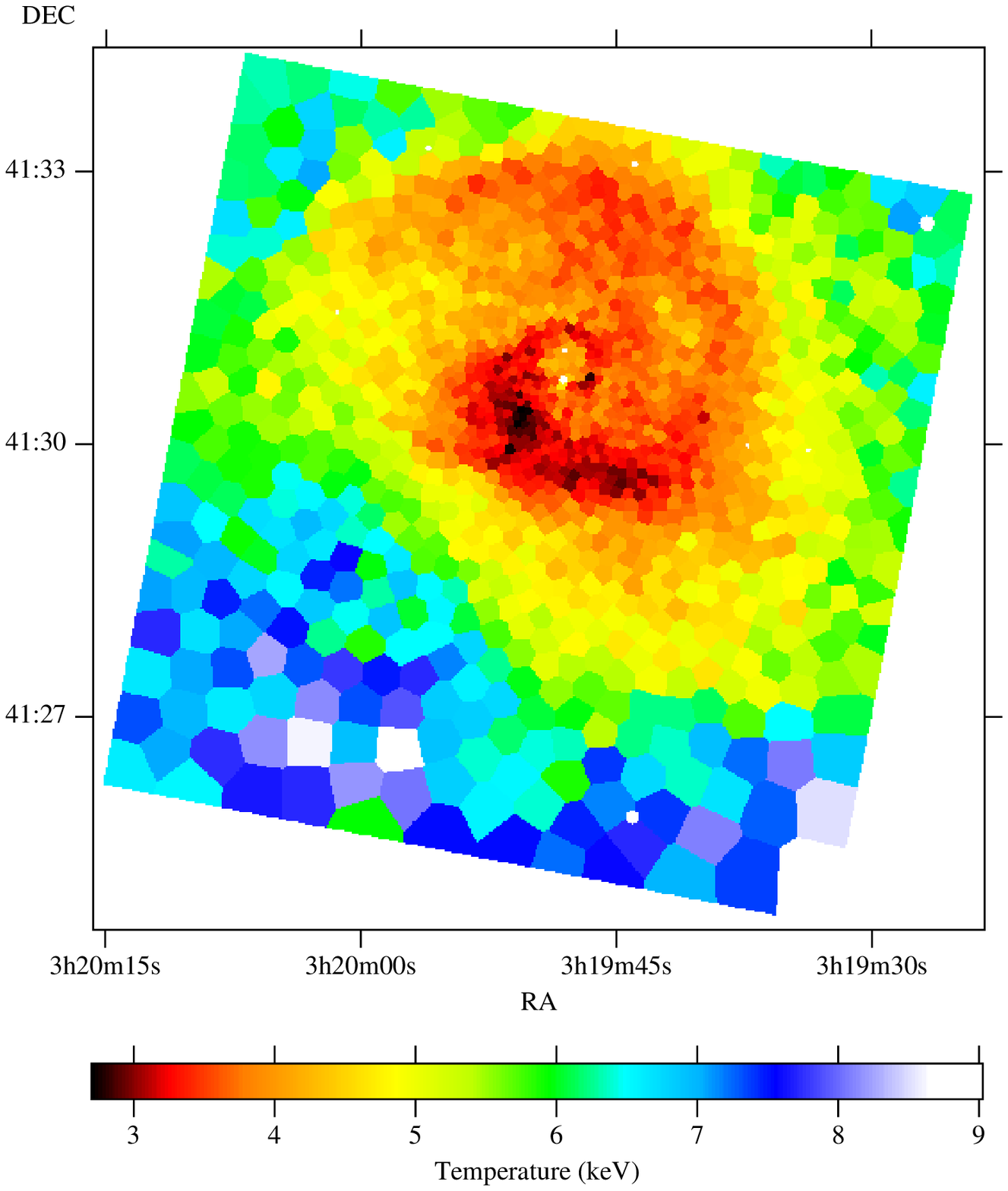}
\includegraphics[width=0.49\columnwidth]{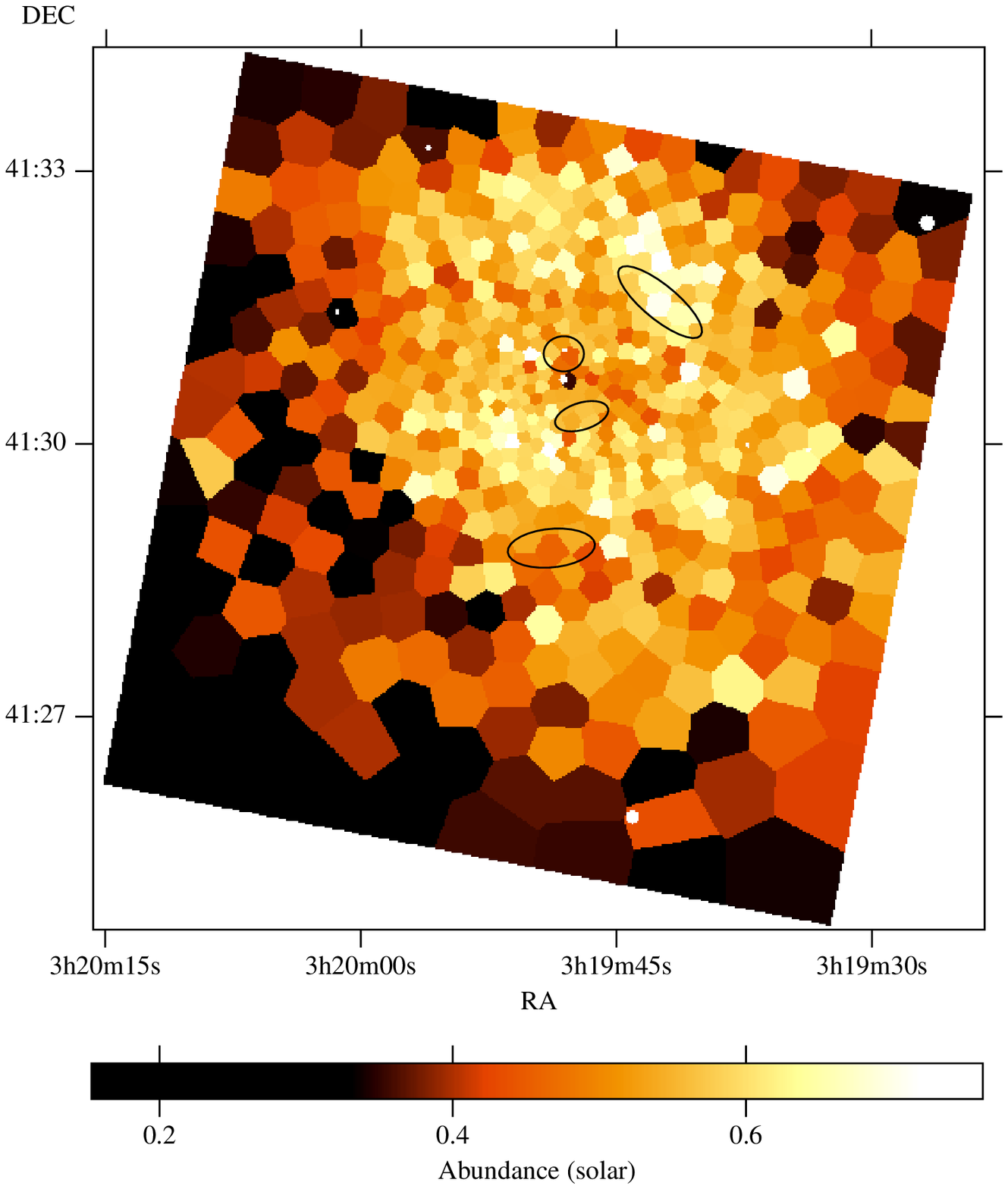}
\caption{Voronoi binning of X-ray data: {\em Left Panel:} Projected emission-weighted temperature map of the core of the Perseus cluster from a 191 ks Chandra observation. {\em Right Panel:} Corresponding map of abundance. (From \protect\cite{2004MNRAS.349..952S}).}
\label{fig:sanders04}
\end{figure}

\begin{figure}[tbp]
\centering
\includegraphics[width=\columnwidth]{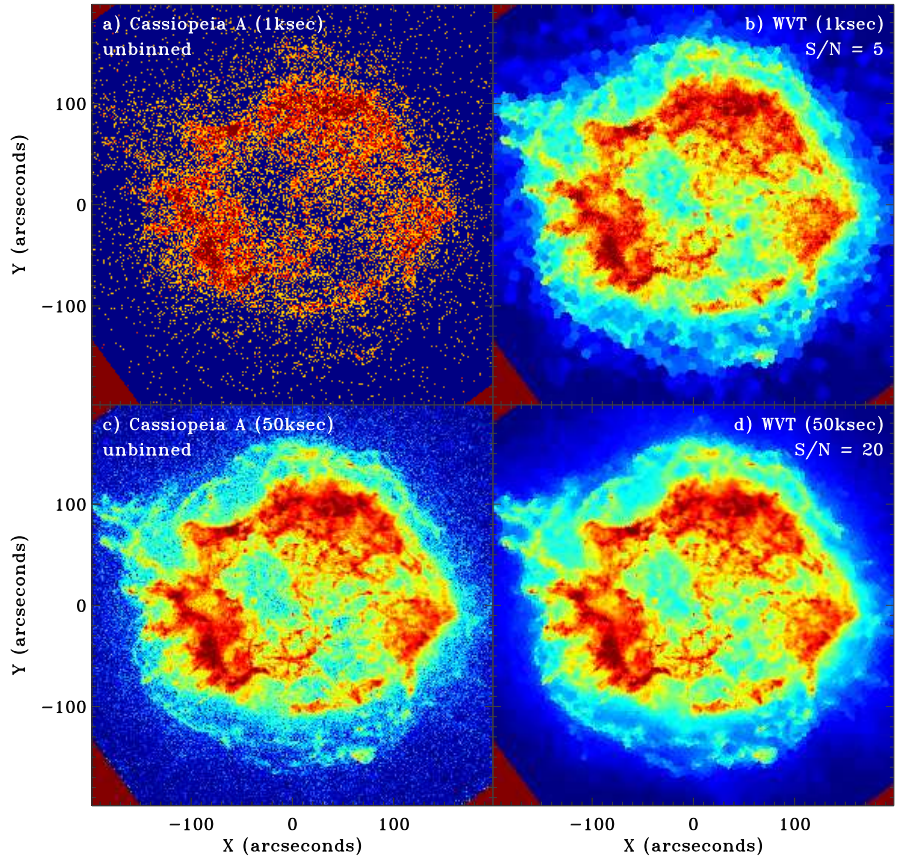}
\caption{Voronoi binning of X-ray data: {\em Top left:} Chandra image of Cassiopeia A with an exposure time of 1 ks. {\em Top right:} the same data adaptively binned to $\snt=5$. {\em Bottom left:} Cassiopeia A with the full exposure of 50 ks. {\em Bottom right:} the same image binned to $\snt=20$. (From \protect\cite{2006MNRAS.368..497D}).}
\label{fig:diehl}
\end{figure}

\section{Applications of the Voronoi binning method}

The Voronoi binning method was introduced and discussed in detail in \cite{2003MNRAS.342..345C}, where it was illustrated on images and integral-field spectroscopic data. The method however was designed to be applicable to general two (or more) dimensional data. In this section we review some representative astrophysical applications of the method since it was introduced.

{\bf X-ray:} The first scientific application of the Voronoi binning method was presented by \cite{2003MNRAS.341..729F}. The authors used the bin-accretion algorithm to produce a colour image from deep Chandra X-ray observation of the cluster environment of radio galaxy 3C~294 at redshift $z\approx1.8$. The bin-accretion method was subsequently used to construct a temperature map from a now classic Chandra X-ray 191 ks observation of the ``shocks and ripples'' at the center of the Perseus cluster \cite{2003MNRAS.344L..43F}. The authors binned the X-ray data-cube by requiring at least 2000 counts per bin ($(\sn)_T\approx45$). The binned X-ray spectra were fitted with models, while varying the temperature, abundance  and column density. A clear visualization of the temperature map from the same data, together with the abundance structure, was also presented by \cite{2004MNRAS.349..952S} and is reproduced in \reffig{fig:sanders04}. It shows that the morphology of the abundance map follows the temperature map closely. An incomplete list of additional  applications of the Voronoi binning method to X-ray data from the Chandra and XMM-Newton satellites includes \cite{2006MNRAS.368..497D,2005MNRAS.360..133S,2005MNRAS.363..216C,2005A&A...444..673S,
2006MNRAS.365..705T,2007ApJ...668..150D,2007A&A...465..749S,
2008ApJ...680..897D,2008A&A...482...97S,2008PASJ...60S..85H,
2008PASJ...60S.173H,2009arXiv0912.0275S,2009A&A...493..409S}. An excellent example of the usefulness of Voronoi binning for the visualization of X-ray data is reproduced in \reffig{fig:diehl}.

\begin{figure}[tbp]
\centering
\includegraphics[width=0.8\columnwidth]{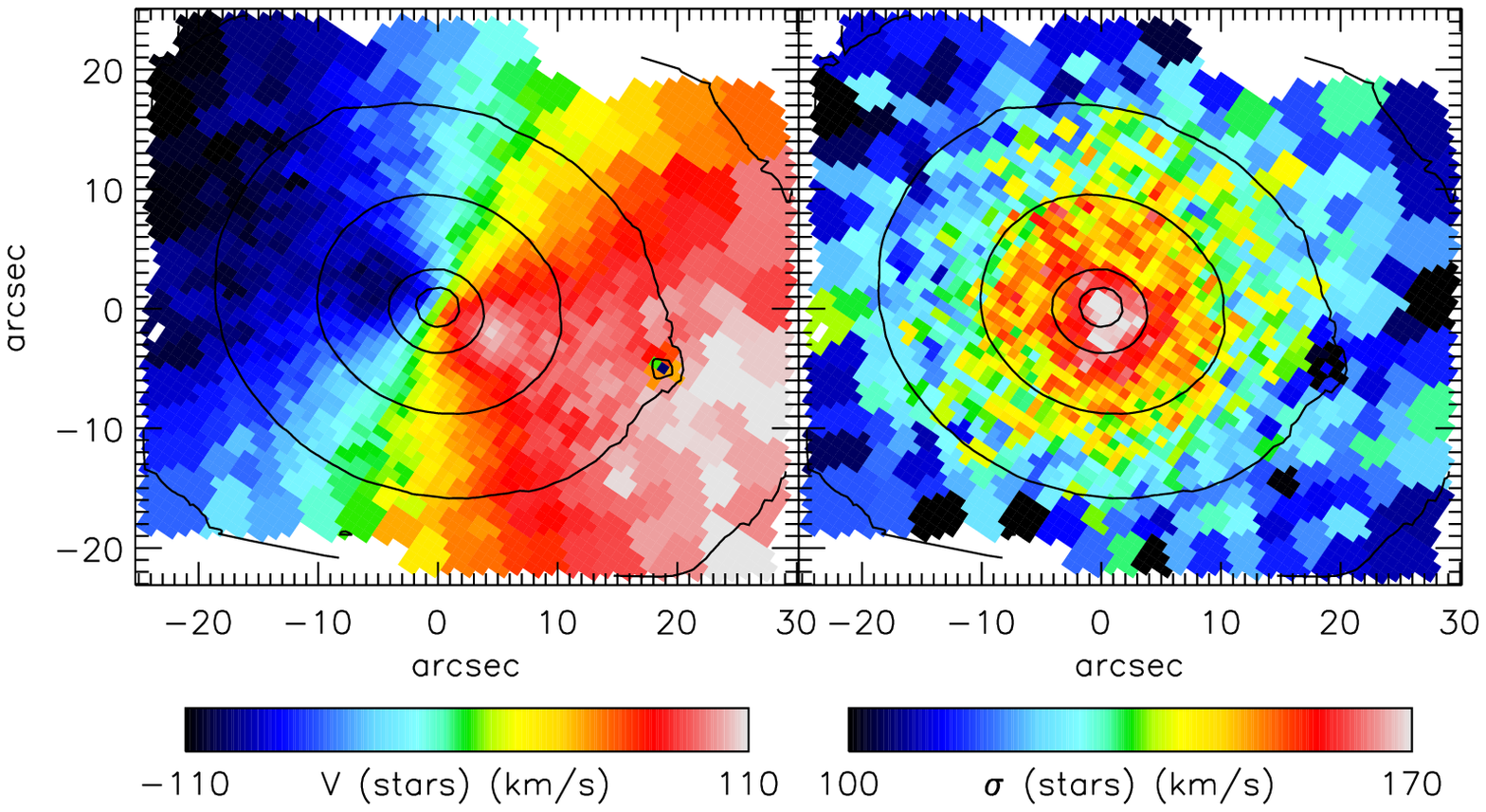}
\includegraphics[width=0.8\columnwidth]{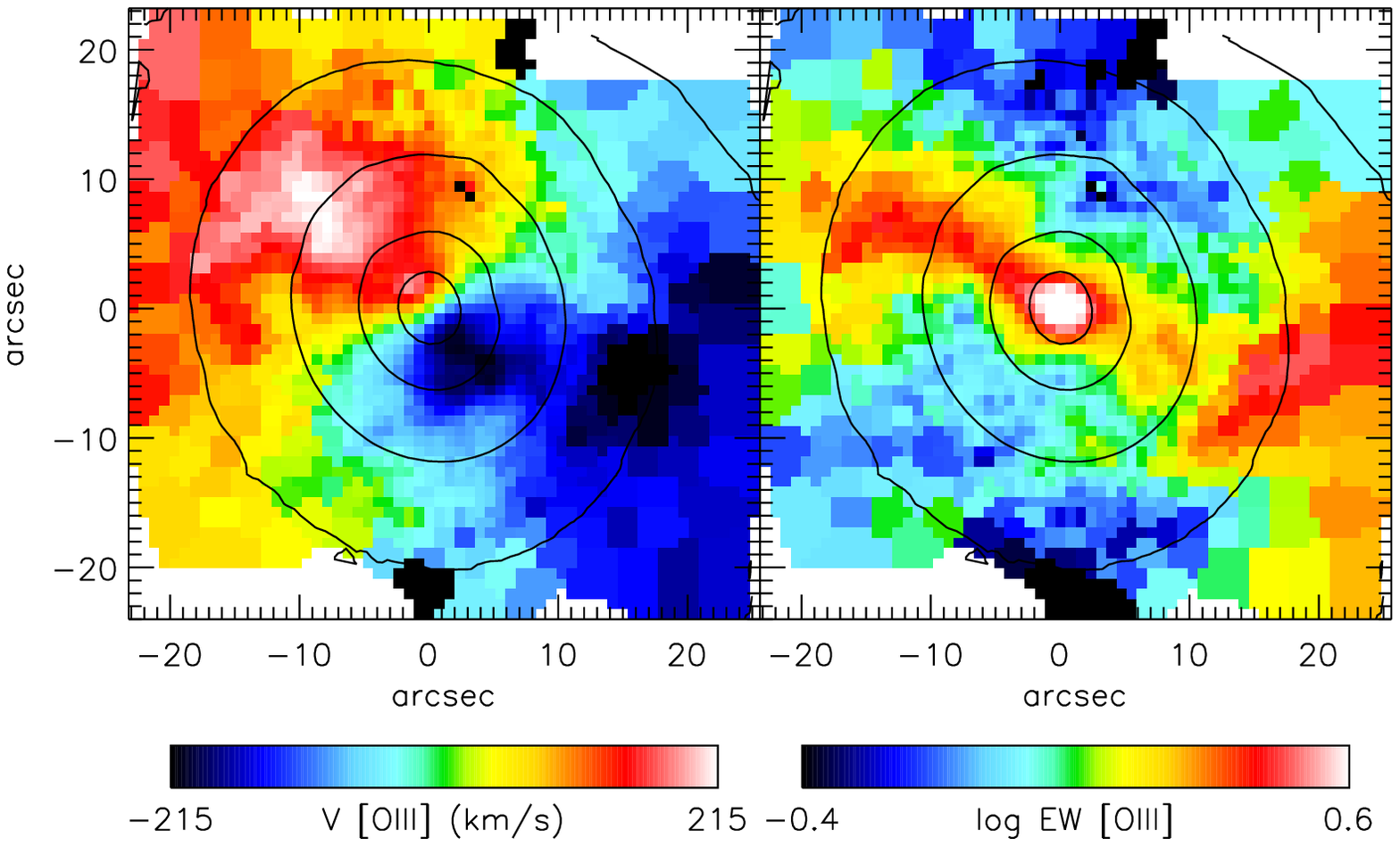}
\caption{Voronoi binning of integral-field spectroscopy: {\em Top Panels:} Mean velocity $V$ and velocity dispersion $\sigma$ of the stars in NGC~4459 (from \protect\cite{2004MNRAS.352..721E}). {\em Bottom Panels:} Mean velocity and equivalent width (EW) of the ionized gas in NGC~4278, as traced by the [OIII] $\lambda\lambda$4959,5007 emission lines (from \protect\cite{2006MNRAS.366.1151S}).}
\label{fig:sauron}
\end{figure}

{\bf Integral-field spectroscopy:} The Voronoi binning method was used to analyze integral-field spectroscopic data from the SAURON survey \cite{2002MNRAS.329..513D} which produced $\sim250,000$ SAURON/OASIS spectra for 72 galaxies. The data were spatially binned to a $(\sn)_T\approx60$, required to reliably extract the stellar kinematics \cite{2004MNRAS.352..721E,2006MNRAS.373..906M}. The same spectra in each bin were then used to also extract the intensity and kinematics of the emission line ionized gas \cite{2006MNRAS.366.1151S,2006MNRAS.369..529F} and the stellar absorptions line-strength \cite{2006MNRAS.369..497K,2007MNRAS.379..445P} (\reffig{fig:sauron}). Some applications by other groups to integral-field data from other spectrographs, include VIMOS \cite{2009ApJ...697L.111C} and SINFONI \cite{2006ApJ...646..754D,2009ApJ...691..749S} on VLT, GMOS \cite{2005AJ....129.2617G,2006MNRAS.365...29G} and NIFS \cite{2008ApJ...687..997S,2009MNRAS.399.1839K} on Gemini, DensePak at WIYN \cite{2006AJ....131..747C}, MPFS on the Russian 6-m SAO \cite{2009MNRAS.394.1229C,2009AstL...35...75S}.

{\bf Dynamical modeling:} The use of binned SAURON data was crucial also to reduce the size of the data in the dynamical modeling of the stellar kinematics for the sample \cite{2006MNRAS.366.1126C,2007MNRAS.379..418C}. Both the computation time and the memory usage in Schwarzschild's orbit-superposition method \cite{1979ApJ...232..236S} scale linearly with the number of data points. The binned data typically had a size $\sim5\times$ smaller than the original data.

\begin{figure}[tbp]
\centering
\includegraphics[width=\columnwidth]{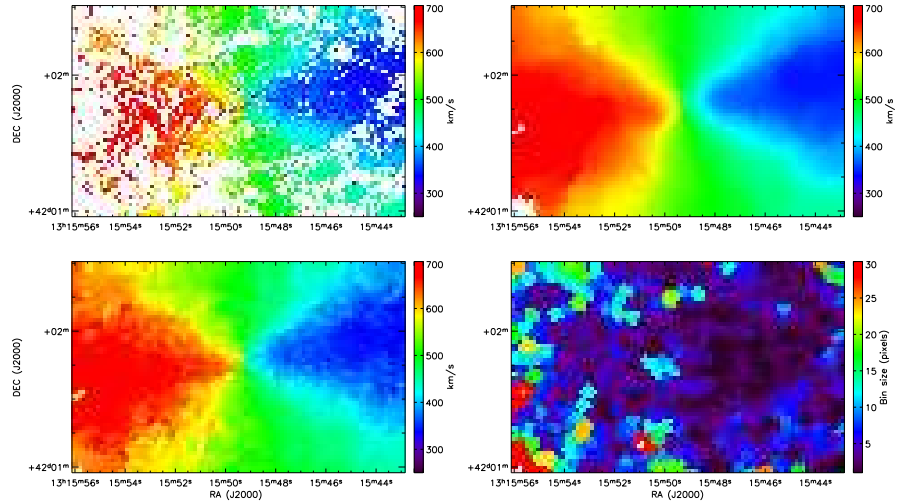}
\caption{Voronoi binning of Fabry-Perot interferometry: {\em Top left:} Unbinned mean gas velocity in the spiral galaxy NGC~5055. {\em Top Right:} Smoothed velocity with a  Gaussian kernel of 6$\times$6 pixels. {\em Bottom left:} Adaptively binned to $\snt=7$. {\em Bottom right:} Bin sizes of the Voronoi binning. (From \protect\cite{2006MNRAS.368.1016D}).}
\label{fig:daigle}
\end{figure}

{\bf Fabry-Perot interferometry:} Voronoi binning was extensively used to extract the H$\alpha$ emission line gas flux and kinematics from Fabry-Perot interferometric observations of spiral galaxies \cite{2005MNRAS.360.1201H,2005ApJ...632..253H,2006MNRAS.366..812C,
2006MNRAS.367..469D,2007A&A...466..905F,2008PASP..120..665H,
2008MNRAS.385..553D,2008AJ....135.2038D,2008MNRAS.388..500E,
2009ApJ...704.1657F}. In this case, due to patchy nature of the observed gas flux, the $\mathcal{S}_j$ could not be reliably pre-computed before the binning. The  \snb\ of the gas was determined by replacing \refeq{eq:sn} with a direct estimation of \snb\ from the spectra during the bin-accretion stage \cite{2006MNRAS.368.1016D}. \reffig{fig:daigle} is a good example of the success of this approach.

\begin{figure}[tbp]
\centering
\includegraphics[width=\columnwidth]{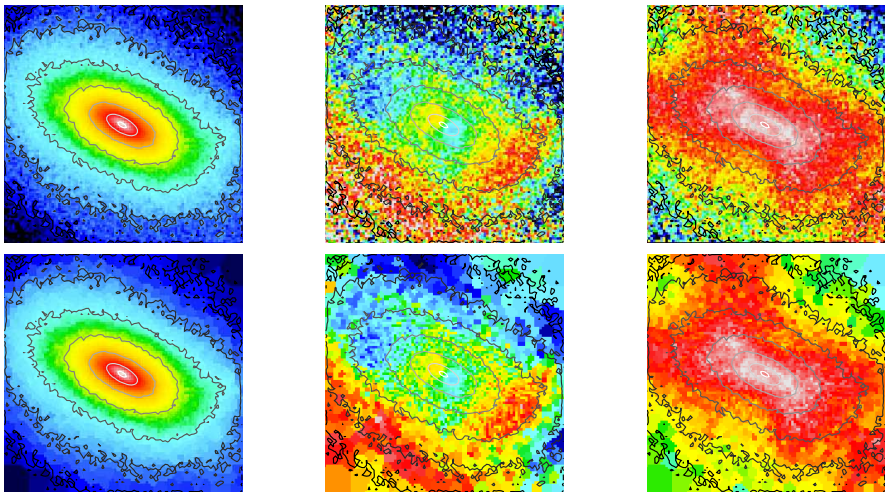}
\caption{Voronoi binning of N-body simulations: {\em Top panels:} From left to right are shown the surface brightness, the mean velocity and the velocity dispersion of a simulated merger remnant. {\em Bottom panels:} Same as in the top panels after adaptive binning to $\snt=10$ (from \protect\cite{Bois2010}).}
\label{fig:bois}
\end{figure}

{\bf N-body simulations:} In the context of the ATLAS$^{3D}$ project\footnote{http://purl.org/atlas3d}, Voronoi binned N-body simulations of galaxy mergers have been used by \cite{Bois2010} to better visualize the kinematics of the remnants in the outer regions and compare them directly with integral-field observations (\reffig{fig:bois}).

\begin{figure}[tbp]
\centering
\includegraphics[width=0.55\columnwidth]{hatch08.ps}
\includegraphics[width=0.43\columnwidth]{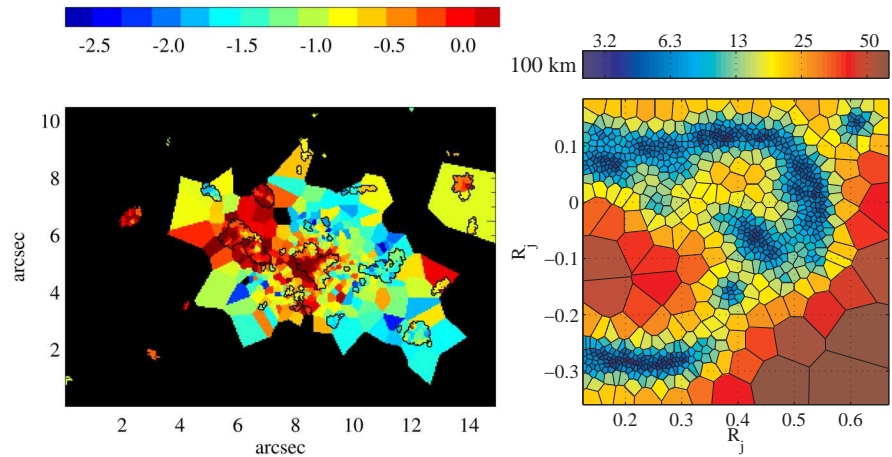}
\caption{Voronoi binning of images: {\em Left panel:} Slope of the UV spectrum of galaxies and intergalactic light of the Spiderweb system (from HST/ACS images). Galaxies are outlined in black. The Ly$\alpha$ flux peaks on the galaxy that lies $\sim3$ arcsec west ($x\approx11''$ arcsec in the figure) of the radio galaxy, which has the bluest colour compared to the other satellite galaxies (from \protect\cite{2008MNRAS.383..931H}). {\em Right panel:} Image extracted from the polar projection of Jupiter's UV auroral emission from an HST/ACS image (from \protect\cite{2009MNRAS.398.1254G}).}
\label{fig:hatch}
\end{figure}

{\bf Standard images:} Photometric studies of galaxies generally adopt ellipse-fitting techniques to analyze radial surface brightness profiles and colours of galaxies (e.g. \cite{1987MNRAS.226..747J}). However in cases where a symmetry along ellipses is not desired, one can Voronoi bin the images to study faint features in the data. Binning was used by \cite{2005ApJ...635..243F,2006ApJ...636..115P,2006MNRAS.370..477L} to study galaxy colours from HST/ACS observations of galaxies at intermediate redshift ($z\sim0.5$). It was employed to look for star formation in the intracluster light of the Spiderweb system (\cite{2008MNRAS.383..931H}, \reffig{fig:hatch}), to study colours of stellar shells in elliptical galaxies merger remnants \cite{2007A&A...467.1011S} and to produce colour maps of early-type galaxies \cite{2008A&A...486...85C,2009A&A...504..389C}. An independent implementation of the Voronoi binning method, which uses a different bin-accretion algorithm, but the same CVT regularization stage, was applied by \cite{2009MNRAS.398.1254G} to study the characteristics of Jupiter's UV auroral emission from HST/ACS images (\reffig{fig:hatch}).

\begin{figure}[tbp]
\hspace{2cm}
\includegraphics[angle=270,width=0.65\columnwidth]{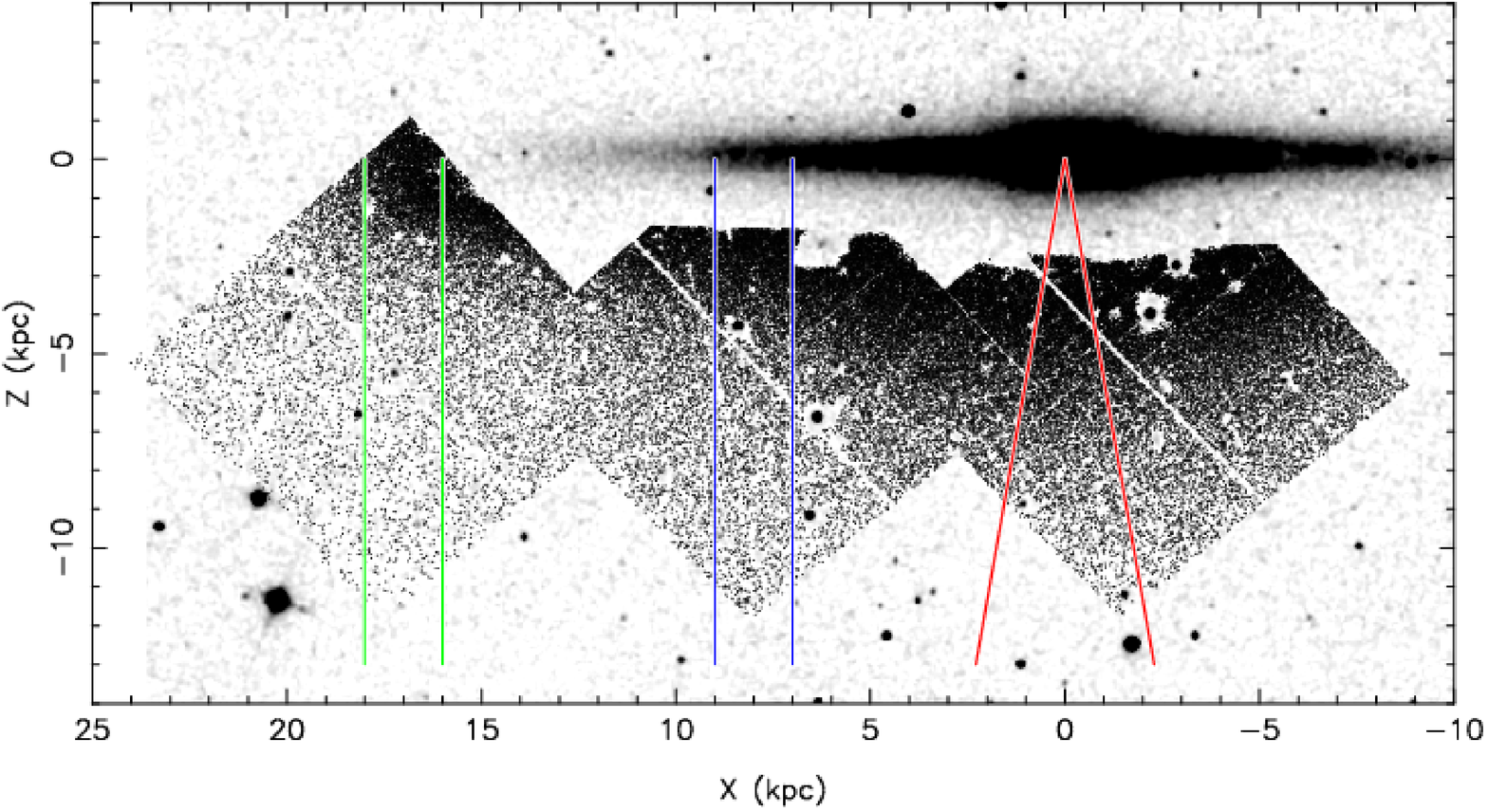}

\hspace{2cm}
\includegraphics[width=0.75\columnwidth]{ibata09b.ps}
\caption{Voronoi binning of star counts: {\em Top panel:} The spatial distribution of point-like sources from three HST/ACS fields is superimposed on the K-band 2MASS image of NGC~891. {\em Bottom panel:} Voronoi binned map of the median metallicity over the ACS survey area. A large-scale gradient in the halo region is noticeable, as well as a more metal-rich "thick disk" (from \protect\cite{2009MNRAS.395..126I}).}
\label{fig:ibata}
\end{figure}

{\bf Star counts:} In \cite{2009MNRAS.395..126I} Voronoi binning was used to group stars counts, as observed with HST/ACS, in the halo of the spiral galaxy NGC~891, an edge-on galaxy considered to be an analogue of the Milky Way. A colour-magnitude diagram was constructed for the stars in each bin to study the variation of the metallicity in the halo. This allowed the authors to trace metallicity gradients at extremely faint surface brightness levels (\reffig{fig:ibata}).

\begin{figure}[tbp]
\centering
\includegraphics[width=0.49\columnwidth]{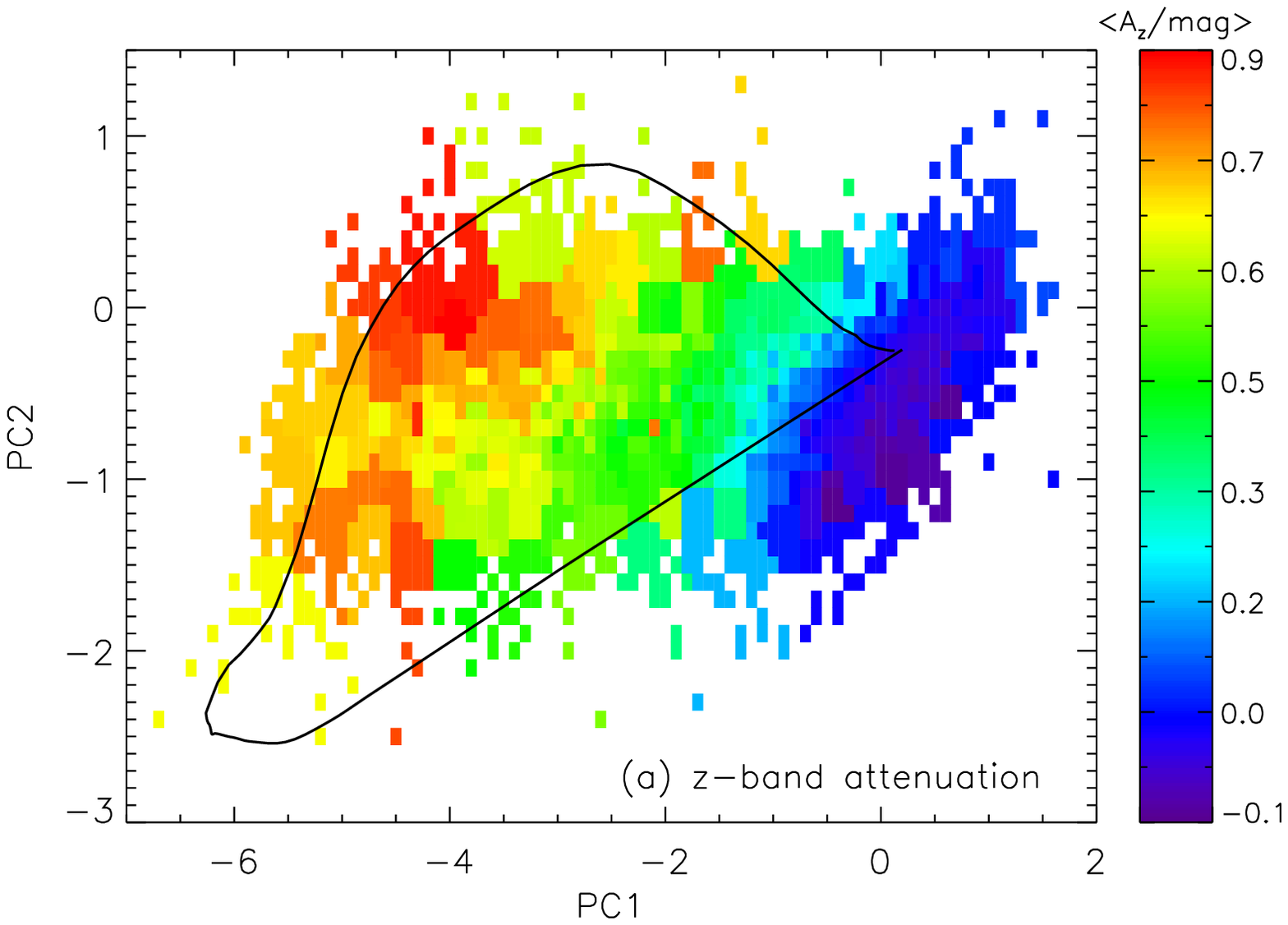}
\includegraphics[width=0.49\columnwidth]{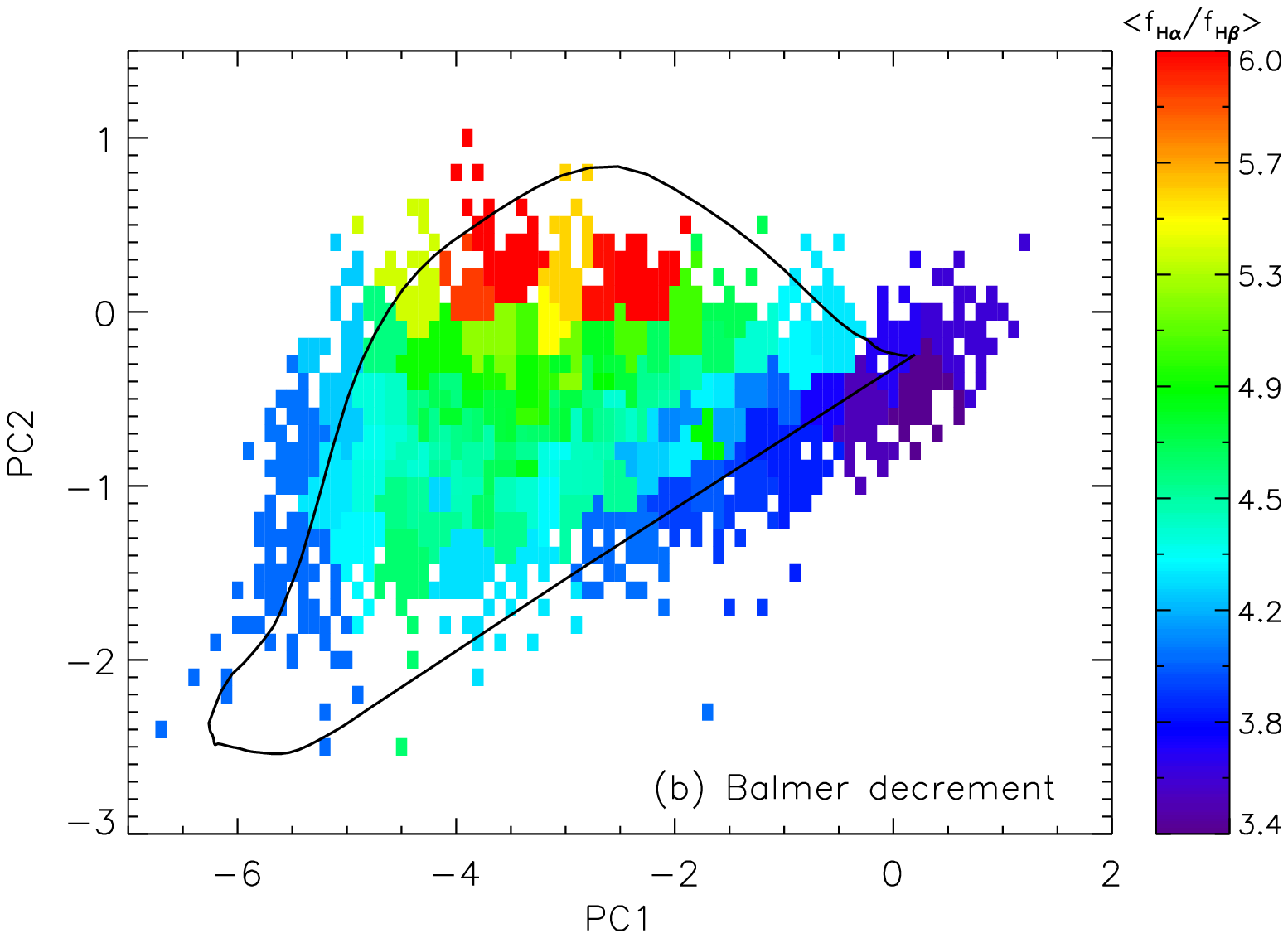}
\caption{Voronoi binning of irregular data: {\em Left panel:} The volume-weighted mean z-band dust attenuation of galaxies in each joint PC1/PC2 histogram bin. {\em Right panel:} The volume-weighted mean Balmer decrement of galaxies in each bin. A single burst track of 1\% burst fraction is overplotted. (From \protect\cite{2007MNRAS.381..543W}).}
\label{fig:wild}
\end{figure}

{\bf General irregular data:} To conclude this overview of applications of the Voronoi binning method, we show in \reffig{fig:wild} the usage of the technique to increase the \sn\ of a generic map of two-dimensional quantities, not arranged in a regular grid. Here the coordinates represent the largest two principal components (PC), as obtained from PC analysis of a large set of spectra from the Sloan Digital Sky Survey. The figure quantifies dust effects as a function of the two PC, which are closely related to the strength of the 4000 \AA\ break and to the equivalent width of the Balmer H$\delta$ absorption line.

\section{Conclusion}

We summarized the motivations and criteria for an optimal binning of two-dimensional data. The ideal binning consists of a Voronoi tessellation described by a lattice of nearly hexagonal bins, which adaptively change their size according to the \sn\ of the data. We described the main concepts of the Voronoi binning method, an algorithm that produces a tessellation satisfying the desired optimal criteria \cite{2003MNRAS.342..345C}.

We reviewed the main applications of the method in astrophysics, in particular to X-ray data, integral-field spectroscopy, Fabry-Perot interferometry, N-body simulations, standard imaging and other general data. The simplicity and generality of the method makes it useful in other fields of science and in the related problem of adaptive mesh generation. No changes to the original concept are required to use the method in three dimensions (e.g. \cite{bustinduy:043901}). We trust that in the future, as it happened in the past, we will see new applications of the technique beyond anything we had anticipated.

\section*{Acknowledgements}

I am grateful to the authors of the plots for their permission to reproduce them in the present review and to Yannick Copin for comments on this paper.


\begin{thebibliography}{10}

\bibitem{2003MNRAS.342..345C}
M.~{Cappellari}, Y.~{Copin}, \mnras \textbf{342}, 345 (2003).

\bibitem{1986desd.book.....S}
B.W. {Silverman}, \emph{{Density estimation for statistics and data analysis}}
  (London: Chapman and Hall, 1986)

\bibitem{1991AJ....102.1581B}
T.C. {Beers}, K.~{Gebhardt}, W.~{Forman}, J.P. {Huchra}, C.~{Jones}, \aj
  \textbf{102}, 1581 (1991).

\bibitem{1994AJ....108..514M}
D.~{Merritt}, B.~{Tremblay}, \aj \textbf{108}, 514 (1994).

\bibitem{1996ApJ...461..622H}
Z.~{Huang}, C.L. {Sarazin}, \apj \textbf{461}, 622 (1996).

\bibitem{1998A&AS..128..397S}
J.~{Starck}, M.~{Pierre}, \aaps \textbf{128}, 397 (1998).

\bibitem{2006MNRAS.368...65E}
H.~{Ebeling}, D.A. {White}, F.V.N. {Rangarajan}, \mnras \textbf{368}, 65
  (2006).

\bibitem{2006MNRAS.368..497D}
S.~{Diehl}, T.S. {Statler}, \mnras \textbf{368}, 497 (2006).

\bibitem{sullivan1991generating}
J.~Sullivan, The Mathematica Journal \textbf{1}(3), 76 (1991)

\bibitem{samet1984quadtree}
H.~Samet, ACM Computing Surveys (CSUR) \textbf{16}(2), 187 (1984)

\bibitem{2001MNRAS.325..178S}
J.S. {Sanders}, A.C. {Fabian}, \mnras \textbf{325}, 178 (2001).

\bibitem{2006MNRAS.371..829S}
J.S. {Sanders}, \mnras \textbf{371}, 829 (2006).

\bibitem{1994MNRAS.270..271V}
R.P. {van der Marel}, \mnras \textbf{270}, 271 (1994)

\bibitem{Okabe2000}
A.~Okabe, B.~Boots, K.~Sugihara, S.~Chiu, \emph{{Spatial tessellations:
  concepts and applications of Voronoi diagrams. 2000}} (Wiley Series in
  Probability and Statistics, 2000)

\bibitem{1986PASP...98.1220R}
J.G. {Robertson}, \pasp \textbf{98}, 1220 (1986).

\bibitem{1986PASP...98..609H}
K.~{Horne}, \pasp \textbf{98}, 609 (1986).

\bibitem{du1999centroidal}
Q.~Du, V.~Faber, M.~Gunzburger, SIAM review \textbf{41}(4), 637 (1999)

\bibitem{lloyd1982least}
S.~Lloyd, IEEE Transactions on Information Theory \textbf{28}(2), 129 (1982)

\bibitem{hartigan1975clustering}
J.~Hartigan, \emph{{Clustering algorithms}} (Wiley New York, 1975)

\bibitem{møller1994lectures}
J.~M{\o}ller, \emph{{Lectures on random Voronoi tessellations}}
  (Springer-Verlag New York, 1994)

\bibitem{du2002grid}
Q.~Du, M.~Gunzburger, Applied mathematics and computation \textbf{133}(2-3),
  591 (2002)

\bibitem{2009MNRAS.tmp.1655S}
V.~{Springel}, \mnras (2009) in press (arXiv:0901.4107).

\bibitem{2004MNRAS.349..952S}
J.S. {Sanders}, A.C. {Fabian}, S.W. {Allen}, R.W. {Schmidt}, \mnras
  \textbf{349}, 952 (2004).

\bibitem{2003MNRAS.341..729F}
A.C. {Fabian}, J.S. {Sanders}, C.S. {Crawford}, S.~{Ettori}, \mnras
  \textbf{341}, 729 (2003).

\bibitem{2003MNRAS.344L..43F}
A.C. {Fabian}, J.S. {Sanders}, S.W. {Allen}, C.S. {Crawford}, K.~{Iwasawa},
  R.M. {Johnstone}, R.W. {Schmidt}, G.B. {Taylor}, \mnras \textbf{344}, L43
  (2003).

\bibitem{2005MNRAS.360..133S}
J.S. {Sanders}, A.C. {Fabian}, R.J.H. {Dunn}, \mnras \textbf{360}, 133 (2005).

\bibitem{2005MNRAS.363..216C}
C.S. {Crawford}, N.A. {Hatch}, A.C. {Fabian}, J.S. {Sanders}, \mnras
  \textbf{363}, 216 (2005).

\bibitem{2005A&A...444..673S}
J.L. {Sauvageot}, E.~{Belsole}, G.W. {Pratt}, \aap \textbf{444}, 673 (2005).

\bibitem{2006MNRAS.365..705T}
G.B. {Taylor}, J.S. {Sanders}, A.C. {Fabian}, S.W. {Allen}, \mnras
  \textbf{365}, 705 (2006).

\bibitem{2007ApJ...668..150D}
S.~{Diehl}, T.S. {Statler}, \apj \textbf{668}, 150 (2007).

\bibitem{2007A&A...465..749S}
A.~{Simionescu}, H.~{B{\"o}hringer}, M.~{Br{\"u}ggen}, A.~{Finoguenov}, \aap
  \textbf{465}, 749 (2007).

\bibitem{2008ApJ...680..897D}
S.~{Diehl}, T.S. {Statler}, \apj \textbf{680}, 897 (2008).

\bibitem{2008A&A...482...97S}
A.~{Simionescu}, N.~{Werner}, A.~{Finoguenov}, H.~{B{\"o}hringer},
  M.~{Br{\"u}ggen}, \aap \textbf{482}, 97 (2008).

\bibitem{2008PASJ...60S..85H}
Y.~{Hyodo}, M.~{Tsujimoto}, K.~{Hamaguchi}, K.~{Koyama}, S.~{Kitamoto},
  Y.~{Maeda}, Y.~{Tsuboi}, Y.~{Ezoe}, \pasj \textbf{60}, 85 (2008)

\bibitem{2008PASJ...60S.173H}
Y.~{Hyodo}, M.~{Tsujimoto}, K.~{Koyama}, S.~{Nishiyama}, T.~{Nagata},
  I.~{Sakon}, H.~{Murakami}, H.~{Matsumoto}, \pasj \textbf{60}, 173 (2008)

\bibitem{2009arXiv0912.0275S}
M.~{Sarzi}, J.C. {Shields}, K.~{Schawinski}, {et al.}, \mnras (2009) in press (arXiv:0912.0275)

\bibitem{2009A&A...493..409S}
A.~{Simionescu}, N.~{Werner}, H.~{B{\"o}hringer}, J.S. {Kaastra},
  A.~{Finoguenov}, M.~{Br{\"u}ggen}, P.E.J. {Nulsen}, \aap \textbf{493}, 409
  (2009).

\bibitem{2004MNRAS.352..721E}
E.~{Emsellem}, M.~{Cappellari}, R.F. {Peletier}, {et al.}, \mnras \textbf{352}, 721 (2004).

\bibitem{2006MNRAS.366.1151S}
M.~{Sarzi}, J.~{Falc{\'o}n-Barroso}, R.L. {Davies}, {et al.}, \mnras
  \textbf{366}, 1151 (2006).

\bibitem{2002MNRAS.329..513D}
P.T. {de Zeeuw}, M.~{Bureau}, E.~{Emsellem}, {et al.}, \mnras \textbf{329}, 513 (2002).

\bibitem{2006MNRAS.373..906M}
R.M. {McDermid}, E.~{Emsellem}, K.L. {Shapiro}, {et al.}, \mnras
  \textbf{373}, 906 (2006).

\bibitem{2006MNRAS.369..529F}
J.~{Falc{\'o}n-Barroso}, R.~{Bacon}, M.~{Bureau}, {et al.}, \mnras
  \textbf{369}, 529 (2006).

\bibitem{2006MNRAS.369..497K}
H.~{Kuntschner}, E.~{Emsellem}, R.~{Bacon}, {et al.}, \mnras \textbf{369}, 497 (2006).

\bibitem{2007MNRAS.379..445P}
R.F. {Peletier}, J.~{Falc{\'o}n-Barroso}, R.~{Bacon}, {et al.}, \mnras
  \textbf{379}, 445 (2007).

\bibitem{2009ApJ...697L.111C}
I.V. {Chilingarian}, S.~{De Rijcke}, P.~{Buyle}, \apjl \textbf{697}, L111
  (2009).

\bibitem{2006ApJ...646..754D}
R.I. {Davies}, J.~{Thomas}, R.~{Genzel}, {et al.}, \apj
  \textbf{646}, 754 (2006).

\bibitem{2009ApJ...691..749S}
F.M. {S{\'a}nchez}, R.I. {Davies}, R.~{Genzel}, L.J. {Tacconi},
  F.~{Eisenhauer}, E.K.S. {Hicks}, S.~{Friedrich}, A.~{Sternberg}, \apj
  \textbf{691}, 749 (2009).

\bibitem{2005AJ....129.2617G}
M.~{Geha}, P.~{Guhathakurta}, R.P. {van der Marel}, \aj \textbf{129}, 2617
  (2005).

\bibitem{2006MNRAS.365...29G}
J.~{Gerssen}, J.~{Allington-Smith}, B.W. {Miller}, J.E.H. {Turner},
  A.~{Walker}, \mnras \textbf{365}, 29 (2006).

\bibitem{2008ApJ...687..997S}
A.C. {Seth}, R.D. {Blum}, N.~{Bastian}, N.~{Caldwell}, V.P. {Debattista}, \apj
  \textbf{687}, 997 (2008).

\bibitem{2009MNRAS.399.1839K}
D.~{Krajnovi{\'c}}, R.M. {McDermid}, M.~{Cappellari}, R.L. {Davies}, \mnras
  \textbf{399}, 1839 (2009).

\bibitem{2006AJ....131..747C}
J.R. {Cort{\'e}s}, J.D.P. {Kenney}, E.~{Hardy}, \aj \textbf{131}, 747 (2006).

\bibitem{2009MNRAS.394.1229C}
I.V. {Chilingarian}, \mnras \textbf{394}, 1229 (2009).

\bibitem{2009AstL...35...75S}
O.K. {Sil'Chenko}, I.V. {Chilingarian}, V.L. {Afanasiev}, Astronomy Letters
  \textbf{35}, 75 (2009).

\bibitem{2006MNRAS.366.1126C}
M.~{Cappellari}, R.~{Bacon}, M.~{Bureau}, {et al.}, \mnras \textbf{366}, 1126 (2006).

\bibitem{2007MNRAS.379..418C}
M.~{Cappellari}, E.~{Emsellem}, R.~{Bacon}, {et al.}, \mnras \textbf{379}, 418 (2007).

\bibitem{1979ApJ...232..236S}
M.~{Schwarzschild}, \apj \textbf{232}, 236 (1979).

\bibitem{2006MNRAS.368.1016D}
O.~{Daigle}, C.~{Carignan}, O.~{Hernandez}, L.~{Chemin}, P.~{Amram}, \mnras
  \textbf{368}, 1016 (2006).

\bibitem{2005MNRAS.360.1201H}
O.~{Hernandez}, C.~{Carignan}, P.~{Amram}, L.~{Chemin}, O.~{Daigle}, \mnras
  \textbf{360}, 1201 (2005).

\bibitem{2005ApJ...632..253H}
O.~{Hernandez}, H.~{Wozniak}, C.~{Carignan}, P.~{Amram}, L.~{Chemin},
  O.~{Daigle}, \apj \textbf{632}, 253 (2005).

\bibitem{2006MNRAS.366..812C}
L.~{Chemin}, C.~{Balkowski}, V.~{Cayatte}, {et al.}, \mnras \textbf{366}, 812 (2006).

\bibitem{2006MNRAS.367..469D}
O.~{Daigle}, C.~{Carignan}, P.~{Amram}, O.~{Hernandez}, L.~{Chemin},
  C.~{Balkowski}, R.~{Kennicutt}, \mnras \textbf{367}, 469 (2006).

\bibitem{2007A&A...466..905F}
K.~{Fathi}, J.E. {Beckman}, A.~{Zurita}, M.~{Rela{\~n}o}, J.H. {Knapen},
  O.~{Daigle}, O.~{Hernandez}, C.~{Carignan}, \aap \textbf{466}, 905 (2007).

\bibitem{2008PASP..120..665H}
O.~{Hernandez}, K.~{Fathi}, C.~{Carignan}, {et al.}, \pasp \textbf{120}, 665 (2008).

\bibitem{2008MNRAS.385..553D}
I.~{Dicaire}, C.~{Carignan}, P.~{Amram}, {et al.}, \mnras \textbf{385}, 553 (2008).

\bibitem{2008AJ....135.2038D}
I.~{Dicaire}, C.~{Carignan}, P.~{Amram}, M.~{Marcelin}, J.~{Hlavacek-Larrondo},
  M.~{de Denus-Baillargeon}, O.~{Daigle}, O.~{Hernandez}, \aj \textbf{135},
  2038 (2008).

\bibitem{2008MNRAS.388..500E}
B.~{Epinat}, P.~{Amram}, M.~{Marcelin}, {et al.}, \mnras
  \textbf{388}, 500 (2008).

\bibitem{2009ApJ...704.1657F}
K.~{Fathi}, J.E. {Beckman}, N.~{Pi{\~n}ol-Ferrer}, O.~{Hernandez},
  I.~{Mart{\'{\i}}nez-Valpuesta}, C.~{Carignan}, \apj \textbf{704}, 1657
  (2009).

\bibitem{Bois2010}
M.~{Bois}, F.~{Bournaud}, E.~{Emsellem}, {et al.}, \mnras (2009) submitted

\bibitem{2008MNRAS.383..931H}
N.A. {Hatch}, R.A. {Overzier}, H.J.A. {R{\"o}ttgering}, J.D. {Kurk}, G.K.
  {Miley}, \mnras \textbf{383}, 931 (2008).

\bibitem{2009MNRAS.398.1254G}
P.~{Guio}, N.~{Achilleos}, \mnras \textbf{398}, 1254 (2009).

\bibitem{1987MNRAS.226..747J}
R.I. {Jedrzejewski}, \mnras \textbf{226}, 747 (1987)

\bibitem{2005ApJ...635..243F}
I.~{Ferreras}, T.~{Lisker}, C.M. {Carollo}, S.J. {Lilly}, B.~{Mobasher}, \apj
  \textbf{635}, 243 (2005).

\bibitem{2006ApJ...636..115P}
A.~{Pasquali}, I.~{Ferreras}, N.~{Panagia}, {et al.}, \apj \textbf{636}, 115 (2006).

\bibitem{2006MNRAS.370..477L}
T.~{Lisker}, V.P. {Debattista}, I.~{Ferreras}, P.~{Erwin}, \mnras \textbf{370},
  477 (2006).

\bibitem{2007A&A...467.1011S}
G.~{Sikkema}, D.~{Carter}, R.F. {Peletier}, M.~{Balcells}, C.~{Del Burgo}, E.A.
  {Valentijn}, \aap \textbf{467}, 1011 (2007).

\bibitem{2008A&A...486...85C}
I.V. {Chilingarian}, V.~{Cayatte}, F.~{Durret}, C.~{Adami}, C.~{Balkowski},
  L.~{Chemin}, T.F. {Lagan{\'a}}, P.~{Prugniel}, \aap \textbf{486}, 85 (2008).

\bibitem{2009A&A...504..389C}
I.V. {Chilingarian}, A.P. {Novikova}, V.~{Cayatte}, F.~{Combes}, P.~{Di
  Matteo}, A.V. {Zasov}, \aap \textbf{504}, 389 (2009).

\bibitem{2009MNRAS.395..126I}
R.~{Ibata}, M.~{Mouhcine}, M.~{Rejkuba}, \mnras \textbf{395}, 126 (2009).

\bibitem{2007MNRAS.381..543W}
V.~{Wild}, G.~{Kauffmann}, T.~{Heckman}, S.~{Charlot}, G.~{Lemson},
  J.~{Brinchmann}, T.~{Reichard}, A.~{Pasquali}, \mnras \textbf{381}, 543
  (2007).

\bibitem{bustinduy:043901}
I.~Bustinduy, F.J. Bermejo, T.G. Perring, G.~Bordel, Review of Scientific
  Instruments \textbf{78}(4), 043901 (2007).

\end{thebibliography}

\end{document}